\documentclass[prl,twocolumn,amsfonts,amsmath,tightenlines,preprintnumbers,nofootinbib,superscriptaddress,letterpaper]{revtex4}
\usepackage{graphicx}
\usepackage{bm}
\usepackage{hyperref}

\newcommand{\lsim}{\raisebox{-0.7ex}{$\stackrel{\textstyle <}{\sim}$ }}

\usepackage{hyperref,amssymb,amsmath,graphicx,xcolor,cancel}

\begin{document}

\preprint{NT@UW-14-19} \preprint{INT-14-038} \preprint{MIT-CTP-4586}
\preprint{RBRC-1090}

\begin{figure}[!t]
  \vskip -1.1cm \leftline{
    \includegraphics[width=3.0 cm]{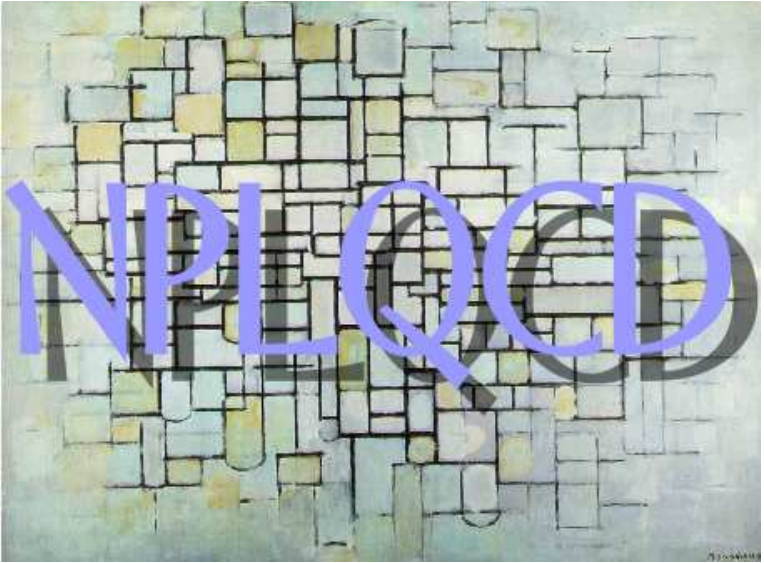}} \vskip
  -0.5cm
\end{figure}

\title{Magnetic moments of light nuclei from lattice quantum
  chromodynamics}

\author{S.R.~Beane} \affiliation{Department of Physics, University of
  Washington, Box 351560, Seattle, WA 98195, USA}
  
\author{E.~Chang} \affiliation{Department of Physics, University of
  Washington, Box 351560, Seattle, WA 98195, USA}
\affiliation{Institute for Nuclear Theory, Box 351550, Seattle, WA
  98195-1550, USA}

\author{S.~Cohen} \affiliation{Department of Physics, University of
  Washington, Box 351560, Seattle, WA 98195, USA}
\affiliation{Institute for Nuclear Theory, Box 351550, Seattle, WA
  98195-1550, USA}

\author{W.~Detmold} \affiliation{ Center for Theoretical Physics,
  Massachusetts Institute of Technology, Cambridge, MA 02139, USA}

\author{H.W.~Lin} \affiliation{Department of Physics, University of
  Washington, Box 351560, Seattle, WA 98195, USA}
  
\author{K.~Orginos} \affiliation{Department of Physics, College of
  William and Mary, Williamsburg, VA 23187-8795, USA}
\affiliation{Jefferson Laboratory, 12000 Jefferson Avenue, Newport
  News, VA 23606, USA}

\author{A.~Parre\~no} \affiliation{Dept. d'Estructura i Constituents
  de la Mat\`eria.  Institut de Ci\`encies del Cosmos (ICC),
  Universitat de Barcelona, Mart\'{\i} Franqu\`es 1, E08028-Spain}

\author{M.J.~Savage} \affiliation{Department of Physics, University of
  Washington, Box 351560, Seattle, WA 98195, USA}
\affiliation{Institute for Nuclear Theory, Box 351550, Seattle, WA
  98195-1550, USA}
 
\author{B.C. Tiburzi} \affiliation{Department of Physics, The City
  College of New York, New York, NY 10031, USA} \affiliation{Graduate
  School and University Center, The City University of New York, New
  York, NY 10016, USA } \affiliation{RIKEN BNL Research Center,
  Brookhaven National Laboratory, Upton, NY 11973, USA }
 
\collaboration{NPLQCD Collaboration}

\date\today

\begin{abstract}
  \noindent
  We present the results of lattice QCD calculations of the magnetic
  moments of the lightest nuclei, the deuteron, the triton and $^3$He,
  along with those of the neutron and proton. These calculations,
  performed at quark masses corresponding to $m_\pi\sim 800~{\rm
    MeV}$, reveal that the structure of these nuclei at unphysically
  heavy quark masses closely resembles that at the physical quark
  masses.  In particular, we find that the magnetic moment of $^3$He
  differs only slightly from that of a free neutron, as is the case in
  nature, indicating that the shell-model configuration of two
  spin-paired protons and a valence neutron captures its dominant
  structure.  Similarly a shell-model-like moment is found for the
  triton, $\mu_{^{3}{\rm H}}\sim \mu_p$. 
  The deuteron magnetic moment
  is found to be equal to the nucleon isoscalar moment within the
  uncertainties of the calculations. Furthermore, deviations from 
  the Schmidt limits are also found to be similar to those in nature for these nuclei. 
  These findings suggest that 
  at least some nuclei at these unphysical quark masses are 
  describable by a phenomenological nuclear shell-model.
\end{abstract}

\maketitle

%%%%%%%%%%%%%%%%%%%%%%%%%%%%%%%%%%%%%%%%%%%%%%%%%%%%%%%%%%%%%%%%
%% Introduction
\noindent
The electromagnetic interactions of nuclei have been used extensively
to elucidate their structure and dynamics.  In the early days of
nuclear physics, the magnetic moments of the light nuclei helped to
reveal that they behaved like a collection of ``weakly'' interacting
nucleons that, to a very large degree, retained their identity,
despite being bound together by the strong nuclear force.  This
feature, in part, led to the establishment of the nuclear shell model
as a phenomenological tool with which to predict basic properties of
nuclei throughout the periodic table.  The success of the shell model
is somewhat remarkable, given that nuclei are fundamentally bound
states of quarks and gluons, the degrees of freedom of quantum
chromodynamics (QCD).  The strong nuclear force emerges from QCD as a
by-product of confinement and chiral symmetry breaking.  The fact
that, at the physical values of the quark masses, nuclei are not
simply collections of quarks and gluons, defined by their global
quantum numbers, but have the structure of interacting protons and
neutrons, remains to be understood at a deep level.  In this letter,
we continue our exploration of nuclei at unphysical quark masses, and
calculate the magnetic moments of the lightest few nuclei using
lattice QCD.  We find that they are close to those found in nature,
and also close to the sum of the constituent nucleon magnetic moments
in the simplest shell model configuration. This second finding in
particular is remarkable and suggests that a phenomenological nuclear
shell-model is applicable for at least some nuclei at these unphysical
quark masses.

Our lattice QCD calculations were performed on one ensemble of
gauge-field configurations generated with a $N_f=3$ clover-improved
fermion action \cite{Sheikholeslami:1985ij} and a L\"uscher-Weisz
gauge action \cite{Luscher:1984xn}. The configurations have $L=32$
lattice sites in each spatial direction, $T=48$ sites in the temporal
direction, and a lattice spacing of $a\sim0.12~{\rm fm}$. All three
light-quark masses were set equal to that of the physical strange
quark, producing a pion of mass $m_\pi \sim 806$ MeV.  A background
electromagnetic ($U_Q(1)$) gauge field giving rise to a uniform
magnetic field along the $z$-axis was multiplied onto each QCD gauge
field in the ensemble (separately for each quark flavor), and these
combined gauge fields were used to calculate up- and down-quark
propagators, which were then contracted to form the requisite nuclear
correlation functions using the techniques of
Ref.~\cite{Detmold:2012eu}.  Calculations were performed on $\sim 750$
gauge-field configurations, taken at uniform intervals from $\sim
10,000$ trajectories.  On each configuration, quark propagators were
generated from 48 uniformly distributed Gaussian-smeared sources for
each of four magnetic field strengths (for further details of the
production, see Refs.~\cite{Beane:2012vq,Beane:2013br}).

Background electromagnetic fields have been used extensively to
calculate electromagnetic properties of single hadrons, such as the
magnetic moments of the lowest-lying
baryons~\cite{Bernard:1982yu,Martinelli:1982cb,Lee:2005ds,
  Lee:2005dq,Detmold:2006vu,Aubin:2008qp,Detmold:2009dx,Detmold:2010ts,Primer:2013pva}
and electromagnetic polarizabilities of mesons and baryons
\cite{Fiebig:1988en,Christensen:2004ca,Lee:2005dq,Detmold:2009dx,Detmold:2010ts,Primer:2013pva,Lujan:2014kia}.
In order that the quark fields, with electric charges $Q_u=+{2\over
  3}$ and $Q_{d,s}=-{1\over 3}$ for the up-, down- and strange-quarks,
respectively, satisfy spatially-periodic boundary conditions in the
presence of a background magnetic field, it is
well-known~\cite{tHooft:1979uj} that the lattice links, $U_\mu(x)$,
associated with the $U_Q(1)$ gauge field are of the form
\begin{eqnarray}
  U_\mu (x)&=& e^{i \frac{6\pi Q_q  \tilde n}{L^2} x_1\delta_{\mu,2}}
  \times e^{-i \frac{6\pi Q_q \tilde n }{L} x_2\delta_{\mu,1}\delta_{x_1,L-1}}
  \ ,
  \label{eq:bkdgfield}
\end{eqnarray}
for quark of flavour $q$, where $\tilde n$ must be an integer.  The uniform magnetic field,
${\bf B}$, resulting from these links is
\begin{equation}
  e {\bf B}  = 
  {6\pi \tilde n \over  L^2}\, \hat {\bf z}
  \ ,
\end{equation}
where $e$ is the magnitude of the electric charge and $\hat{\bf z}$ is
a unit vector in the $x_3$-direction.  In physical units, the
background magnetic fields exploited with this ensemble of gauge-field
configurations are $e |{\bf B}| \sim 0.046 \ |\tilde n|\ {\rm GeV}^2$.
To optimize the re-use of light-quark propagators in the production,
calculations were performed for $U_Q(1)$ fields with $\tilde
n=0,1,-2,+4$.  Four field strengths were found to be sufficient for
this initial investigation.  With three degenerate flavors of light
quarks, and a traceless electric-charge matrix, there are no
contributions from coupling of the ${\bf B}$ field to sea quarks at
leading order in the electric charge.  Therefore, the 
 magnetic moments presented here are complete calculations (there are 
 no missing disconnected contributions).

The ground-state energy of a non-relativistic hadron of mass $M$, and 
charge $Q\, e$  in a uniform magnetic field is
\begin{eqnarray}
  \!\!\!\!\!\!\!E({\bf B}) & = & M\ +\ {| Q\, e\, {\bf B}|\over 2 M}
  \ -\ {\bm\mu}\cdot {\bf B} 
  \nonumber \\ &&
  \ -\ 2\pi \beta_{M0}\ |{\bf B}|^2 
  -2\pi \beta_{M2} T_{ij} B_i B_j+ ... \ ,
  \label{eq:energyB}
\end{eqnarray}
where the ellipses denote terms that are cubic and higher in the
magnetic field, as well as terms that are $1/M$
suppressed~\cite{Hill:2011wy,Lee:2014iha}.  The first contribution in
eq.~(\ref{eq:energyB}) is the hadron's rest mass, the second is the
energy of the lowest-lying Landau level, the third is from the
interaction of its magnetic moment, ${\bm\mu}$, and the fourth and
fifth terms are from its scalar and quadrupole magnetic
polarizabilities, $\beta_{M0,M2}$, respectively 
($T_{ij}$ is a 
traceless symmetric tensor \cite{Chen:1998vi}).  The magnetic moment
term is only present for particles with spin, and $\beta_{M2}$ is only
present for $j\geq1$.  In order to determine ${\bm\mu}$ using lattice
QCD calculations, two-point correlation functions associated with the
hadron or nucleus of interest in the $j_z=\pm j$ magnetic sub-states,
$C^{(B)}_{j_z}(t)$, can be calculated in the presence of background
fields of the form given in Eq.~(\ref{eq:bkdgfield}) with strength
$B=\hat{\bf z}\cdot{\bf B}$.  The energies of ground-states aligned
and anti-aligned with the magnetic field, $E^B_{\pm j}$, will be split
by spin-dependent interactions, and the difference, $\delta E^{(B)} =
E^{B}_{+j} - E^{B}_{-j}$, can be extracted from the correlation
functions that we consider.  The component of $\delta E^{(B)}$ that is
linear in ${\bf B}$ determines ${\bm\mu}$ via Eq.~(\ref{eq:energyB}).
\begin{figure}[!t]
  \centering
  \includegraphics[width=0.99\columnwidth]{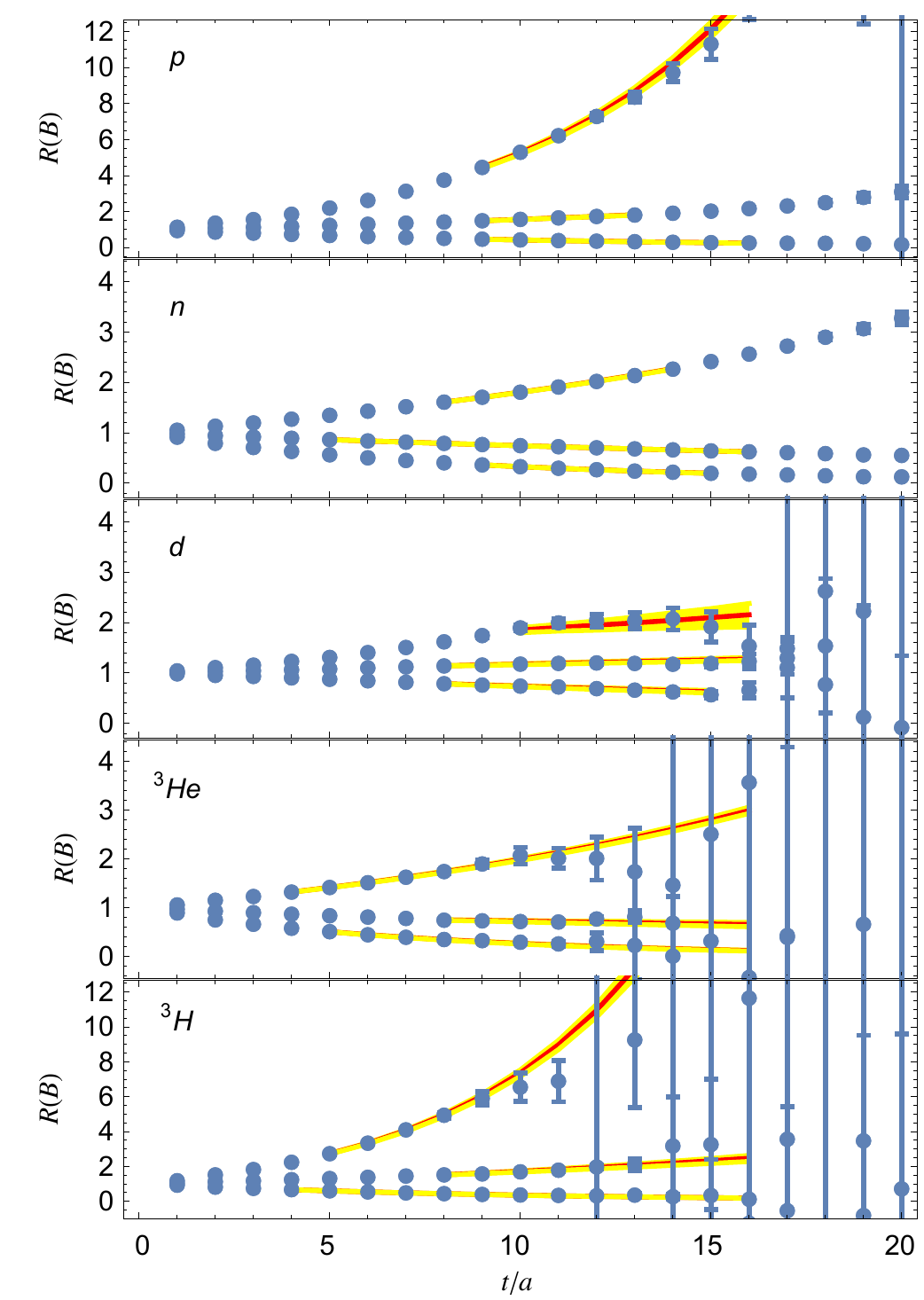}\ \
  \caption{ The correlator ratios $R(B)$ as a function of time slice
    for the various states ($p$, $n$, $d$, ${}^3$He, and ${}^3$H) for
    $\tilde n=+1,-2,+4$.  Fits to the ratios are also shown.}
  \label{fig:EMP}
\end{figure}
Explicitly, the energy difference is determined from the large time
behaviour of
\begin{eqnarray}
  R(B) =
  \frac{C^{(B)}_{j}(t)\ C^{(0)}_{-j}(t)}{ C^{(B)}_{-j}(t)\ C^{(0)}_{j}(t)}
  &\stackrel{t\to\infty}{\longrightarrow} &Z e^{-\delta E^{(B)} t}
  \,.
  \label{eq:ratio}
\end{eqnarray}
Each term in this ratio is a correlation function with the quantum
numbers of the nuclear state that is being considered, which we
compute using the methods of Ref.~\cite{Detmold:2012eu}.  As discussed
in Ref.~\cite{Primer:2013pva}, subtracting the contribution from the
correlation functions calculated in the absence of a magnetic field
reduces fluctuations in the ratio, enabling a more precise
determination of the magnetic moment.  The energy splitting is
extracted from a correlated $\chi^2$-minimization of the functional
form in Eq.~(\ref{eq:ratio}) using a covariance matrix generated with
the jackknife procedure. Fits are performed only over time ranges
where all of the individual correlators in the ratio exhibit single
exponential behavior and a systematic uncertainty is assigned from
variation of the fitting window.  Figure~\ref{fig:EMP} shows the
correlator ratios and associated fits for the various states that we
consider: $p$, $n$, $d$, ${}^3$He, and ${}^3$H, for $\tilde
n=+1,-2,+4$.

As mentioned above, the magnetic moments of the proton and neutron
have been previously calculated with lattice QCD methods for a wide
range of light-quark masses (in almost all cases omitting the
disconnected contributions).  The present work is the first QCD
calculation of the magnetic moments of nuclei.
\begin{figure}[!t]
  \centering
  \includegraphics[width=0.99\columnwidth]{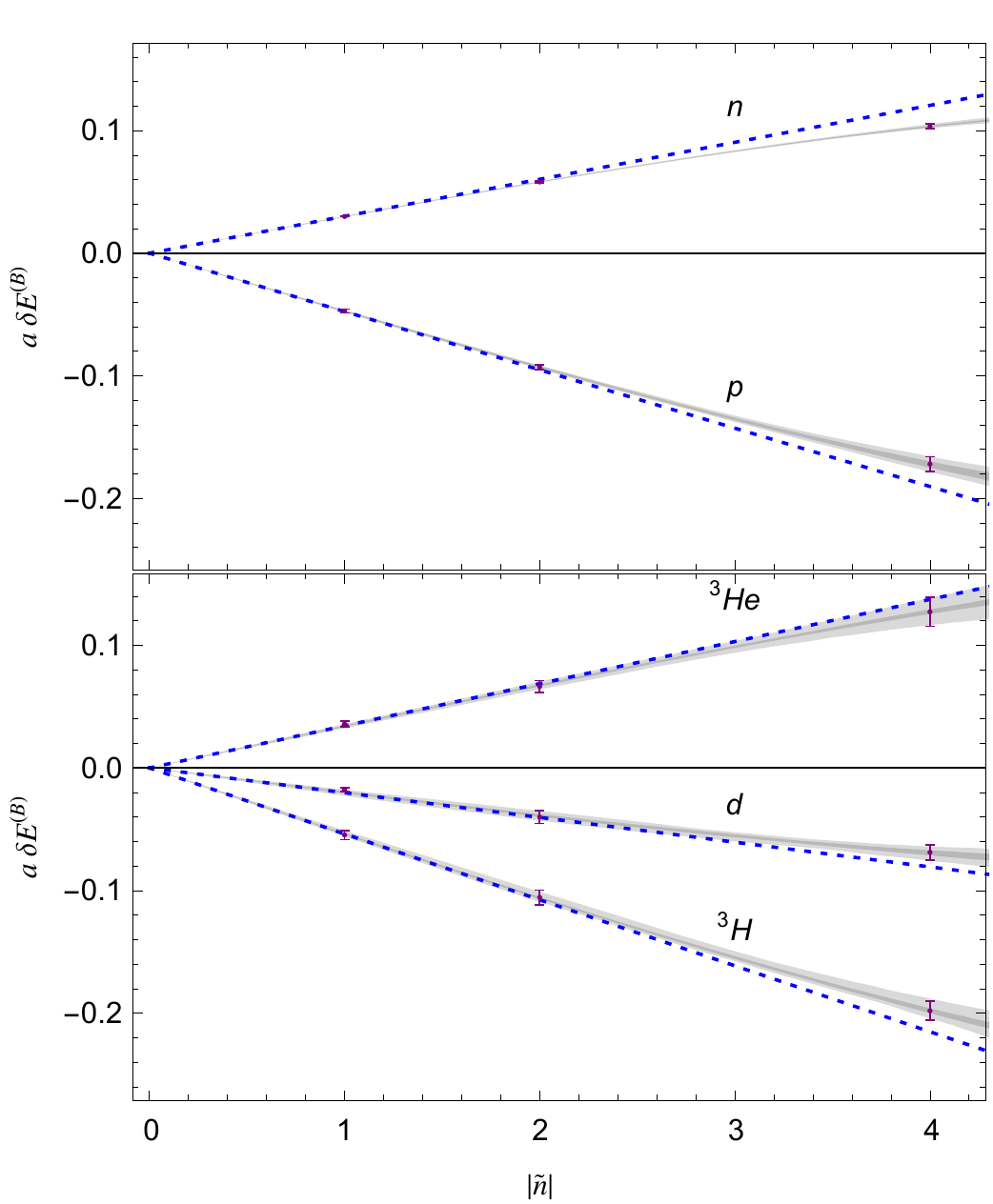}\ \
  \caption{ The calculated $\delta E^{(B)}$ of the proton and neutron
    (upper panel) and light nuclei (lower panel) in lattice units as a
    function of $|\tilde n|$.  The shaded regions corresponds to fits
    of the form $\delta E^{(B)} =-2 \mu\ |{\bf B}| + \gamma\, |{\bf
      B}|^3 $ and their uncertainties.  The dashed lines correspond to
    the linear contribution alone.  }
  \label{fig:dEnp}
\end{figure}
In Figure~\ref{fig:dEnp}, we show the energy splittings of the
nucleons and nuclei as a function of $|\tilde n|$, and, motivated by
Eq.~(\ref{eq:energyB}), we fit these to a function of the form $\delta
E^{(B)} =-2 \mu\ |{\bf B}| + \gamma\ |{\bf B}|^3 $, where $\gamma$ is
a constant encapsulating higher-order terms in the expansion.  We find
that the proton and neutron magnetic moments at this pion mass are
$\mu_p = 1.792(19)(37)$ NM (nuclear magnetons) and $\mu_n =
-1.138(03)(10)$ NM, respectively, where the first uncertainty is
statistical and the second uncertainty is from systematics associated
with the fits to correlation functions and the extraction of the
magnetic moment using the above form.  These results agree with
previous calculations \cite{Primer:2013pva} within the uncertainties.
The natural units of the system are ${e/
  2 M_N^{\rm latt}}$, where $M_N^{\rm latt}$ is the mass of the nucleon at the quark masses
of the lattice calculation, which we refer to as natural nuclear magnetons (nNM). 
In these units, the magnetic moments are $\mu_p =
3.119(33)(64)$ nNM and $\mu_n = -1.981(05)(18)$ nNM.  These values at
this unphysical pion mass can be compared with those of nature,
$\mu_p^{\rm expt} = 2.792847356(23)$ NM and $\mu_n^{\rm expt} =
-1.9130427(05)$ NM, which are remarkably close to the lattice results.
In fact, when comparing all available lattice QCD results for the
nucleon magnetic moments in units of nNM, the dependence upon the
light-quark masses is surprisingly small, reminiscent of the almost
completely flat pion mass dependence of the nucleon axial coupling,
$g_A$.

In Figure~\ref{fig:dEnp}, we also show $\delta E^{(B)}$ as a function
of $|\tilde n|$ for the deuteron, $^3$He and the triton (${}^3$H).
Fitting the energy splittings with a form analogous to that for the
nucleons gives magnetic moments of $\mu_d = 1.218(38)(87)$ nNM for the
deuteron, $\mu_{^3{\rm He}} = -2.29(03)(12)$ nNM for $^3$He and
$\mu_{^3{\rm H}} = 3.56(05)(18)$ nNM for the triton.  These can be
compared with the experimental values of $\mu_d^{\rm expt} =
0.8574382308(72)$ NM, $\mu_{^3{\rm He}}^{\rm expt} = -2.127625306(25)$
NM and $\mu_{^3{\rm H}}^{\rm expt} = 2.978962448(38)$ NM.
\begin{figure}[!t]
  \centering
  \includegraphics[width=0.99\columnwidth]{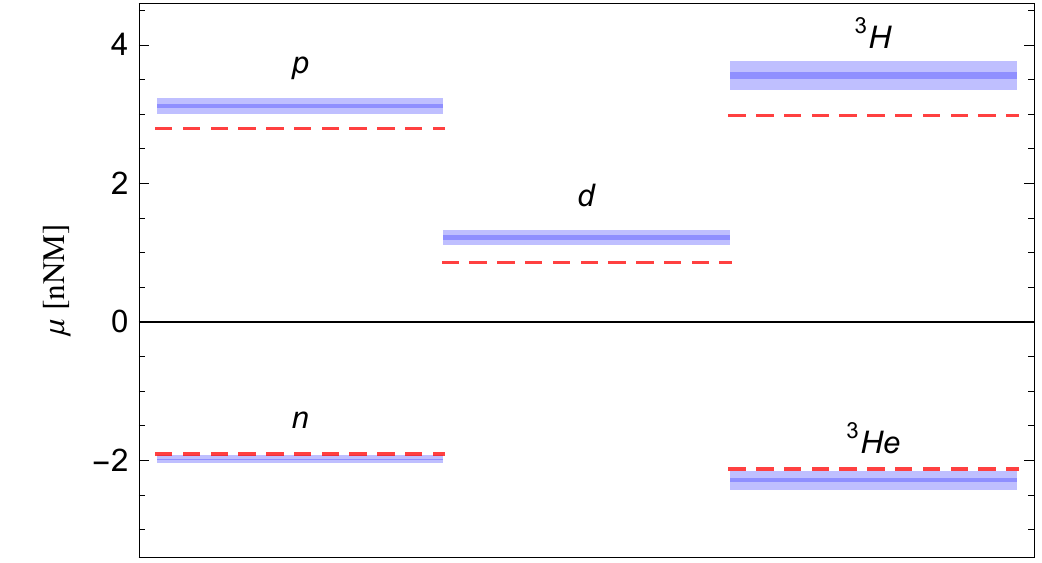}\
  \
  \caption{ The magnetic moments of the proton, neutron, deuteron,
    $^3$He and triton.  The results of the lattice QCD calculation at
    a pion mass of $m_\pi\sim 806~{\rm MeV}$, in units of natural
    nuclear magnetons (${e}/{2M_N^{\rm latt}}$), are shown as the solid bands.  The inner bands
    corresponds to the statistical uncertainties, while the outer
    bands correspond to the statistical and systematic uncertainties
    combined in quadrature, and include our estimates of the
    uncertainties from lattice spacing and volume.  The red dashed
    lines show the experimentally measured values at the physical
    quark masses.  }
  \label{fig:Nuclei}
\end{figure}
The magnetic moments calculated with lattice QCD, along with their
experimental values, are presented in Figure~\ref{fig:Nuclei}.  The
naive shell-model predictions for the magnetic moments of these light nuclei
are $\mu_d^{\rm SM} = \mu_p+\mu_n$, $\mu^{\rm SM}_{^3{\rm He}} =
\mu_n$ (where the two protons in the 1s-state are spin paired to
$j_p=0$ and the neutron is in the 1s-state) and $\mu^{\rm SM}_{^3{\rm
    H}} = \mu_p$ (where the two neutrons in the 1s-state are spin
paired to $j_n=0$ and the proton is in the 1s-state).  For these
simple s-shell nuclei, the proton and neutron magnetic moments
correspond to the Schmidt limits \cite{schmidt:1937}.  In nature, $^3$He is one of the
very few nuclei that lie outside the Schmidt limits \cite{foldy:1950}.  In our
calculations we also find that $^3$He lies outside the Schmidt limits
at this heavier pion mass, with $\delta\mu_{^3{\rm He}} = \mu_{^3{\rm
    He}} - \mu_n = -0.340(24)(93)$ nNM (compared to the experimental
difference of $\delta\mu_{^3{\rm He}}^{\rm expt} = -0.215$ NM) , and
similarly for the triton $\delta\mu_{^3{\rm H}}=\mu_{^3{\rm H}} -
\mu_p = +0.45(04)(16)$ nNM (compared to the experimental difference of
$\delta\mu_{^3{\rm H}}^{\rm expt} = +0.186$ NM), corresponding to
$\sim 10\%$ deviations from the naive shell-model predictions.
\begin{figure}[!t]
  \centering
  \includegraphics[width=0.99\columnwidth]{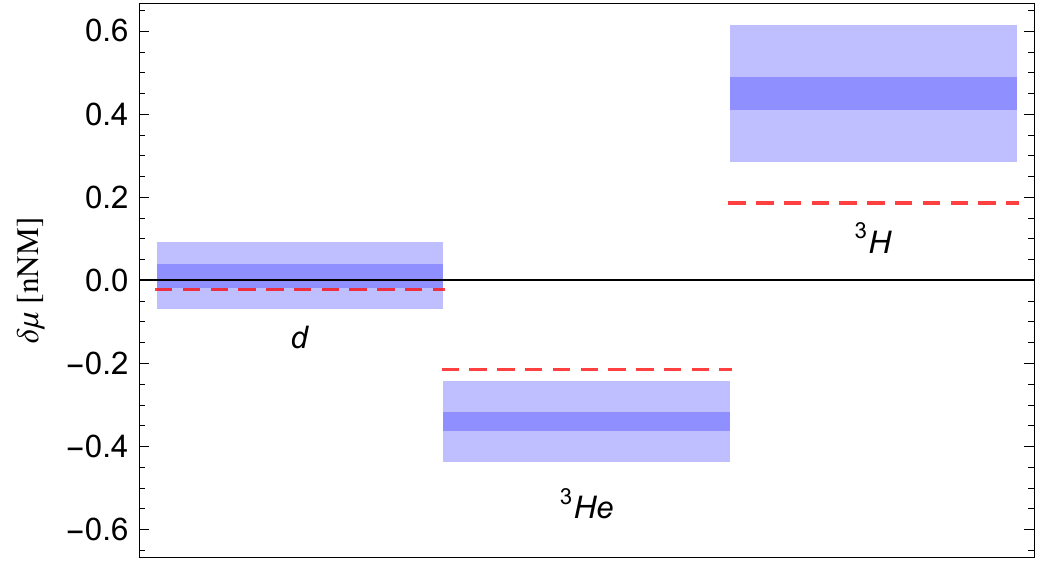}\
  \
  \caption{ The differences between the nuclear magnetic moments and
    the predictions of the naive shell-model.  The results of the
    lattice QCD calculation at a pion mass of $m_\pi\sim 806~{\rm
      MeV}$, in units of natural nuclear magnetons (${e}/{2M_N^{\rm latt}}$), are shown as the
    solid bands. The inner band corresponds to the statistical
    uncertainties, while the outer bands correspond to the statistical
    and systematic uncertainties combined in quadrature, including
    estimates of the uncertainties from lattice spacing and volume.
    The red dashed lines show the experimentally measured differences.
  }
  \label{fig:diffNuclei}
\end{figure}
These quantities are summarized in Figure~\ref{fig:diffNuclei}.

At a phenomenological level, it is not difficult to understand why the
magnetic moments scale, to a large degree, with the nucleon mass.  The
success of the non-relativistic quark model (NRQM) in describing the
magnetic moments of the lowest-lying baryons as the sum of
contributions from three weakly-bound non-relativistic quarks, with
up- and down-quark masses of $M_{U,D}\sim 300~{\rm MeV}$ and
strange-quark mass of $M_{S}\sim 500~{\rm MeV}$, suggests that naive
scaling with the hadron mass should capture most of the quark-mass
dependence.  From the perspective of chiral perturbation theory
($\chi$PT), the leading contributions to the nucleon magnetic moments
are from dimension-five operators, with the leading quark-mass
dependence arising from mesons loops that are suppressed in the chiral
expansion, and scaling linearly with the mass of the pion.
Consistency of the magnetic moments calculated in the NRQM and in
$\chi$PT suggests that the nucleon mass scales linearly with the pion
mass, which is inconsistent with chiral power counting, but consistent
with the results obtained from analysis of lattice QCD
calculations~\cite{Walker-Loud:2013yua}. It should be emphasized that
the magnetic moments of the light nuclei that we study here are well
understood in the context of nuclear chiral effective field theory,
where pions and nucleons are the effective degrees of freedom, and
heavier meson-exchange-type contributions are included as various
contact interactions among nucleons (see, for instance,
Ref.~\cite{Piarulli:2012bn}).

The present calculations have been performed at a single lattice
spacing and in one lattice volume, and the lack of continuum and
infinite volume extrapolations introduces systematic uncertainties
into our results. Chiral perturbation theory can be used to estimate
the finite volume (FV) effects in the magnetic moments, using the sum
of the known~\cite{Tiburzi:2014zva} effects on the constituent
nucleons. These contributions are $\lsim 1\%$ in all cases.  There may
be additional effects beyond the single particle contributions,
however the binding energies of light nuclei calculated previously in
multiple volumes at this quark mass \cite{Beane:2012vq} demonstrate
that the current lattice volume is large enough for such FV effects to
be negligible. In contrast, calculations with multiple lattice
spacings have not been performed at this heavier pion mass, and
consequently this systematic uncertainty remains to be quantified.
However, electromagnetic contributions to the action are
perturbatively improved as they are included as a background field in
the link variables. Consequently, the lattice spacing artifacts are
expected to be small, entering at ${\cal O}(\Lambda_{\rm
  QCD}^2a^2)\sim 3$\% for $\Lambda_{\rm QCD}=300$ MeV. To account for
these effects, we combine the two sources of uncertainty in quadrature
and assess an overall multiplicative systematic uncertainty of 3\% on
all the extracted moments. For the nuclei, this is small compared to
the other systematic uncertainties, but for the neutron in particular,
it is the dominant uncertainty.

In conclusion, we have presented the results of lattice QCD
calculations of the magnetic moments of the lightest nuclei at the
flavor SU(3) symmetric point.  We find that, when rescaled by the mass
of the nucleon, the magnetic moments of the proton, neutron, deuteron,
$^3$He and triton are remarkably close to their experimental values.
The magnetic moment of $^3$He is very close to that of a free neutron,
consistent with the two protons in the 1s-state spin-paired to $j_p=0$
and the valence neutron in the 1s-state.  Analogous results are found
for the triton, and the magnetic moment of the deuteron is consistent
with the sum of the neutron and proton magnetic moments.  This work
demonstrates for the first time that QCD can be used to calculate the
structure of nuclei from first principles. Calculations using these
techniques at lighter quark masses and for larger nuclei are ongoing
and will be reported in future work. Perhaps even more importantly,
these results reveal aspects of the nature of nuclei, not at the
physical quark masses, but in a more general setting where Standard
Model parameters are allowed to vary.  In particular, they indicate
that the phenomenological successes of the nuclear shell-model in
nature may extend over a broad range of quark masses.

\vskip 0.25in

\noindent 
We thank D.B.~Kaplan and D.R.~Phillips for helpful discussions.  SRB
was supported in part by NSF continuing grant PHY1206498, MJS was
supported in part by DOE grant No.~DE-FG02-00ER41132, WD was supported
by the U.S. Department of Energy Early Career Research Award
DE-SC0010495 and the Solomon Buchsbaum Fund at MIT. KO was supported
by the U.S. Department of Energy through Grant Number DE-
FG02-04ER41302 and through Grant Number DE-AC05-06OR23177 under which
JSA operates the Thomas Jefferson National Accelerator Facility.  HWL
was supported by DOE grant No.~DE-FG02-97ER4014.  The work of AP was
supported by the contract FIS2011-24154 from MEC (Spain) and FEDER.
BCT was supported in part by a joint City College of New
York--RIKEN/Brookhaven Research Center fellowship, a grant from the
Professional Staff Congress of the CUNY, and by the U.S.~National
Science Foundation, under Grant No.~PHY$12$-$05778$.  This work made
use of high-performance computing resources provided by XSEDE
(supported by National Science Foundation Grant Number OCI-1053575),
NERSC (supported by U.S. Department of Energy Grant Number
DE-AC02-05CH11231), the PRACE Research Infrastructure resource Mare
Nostrum at the Barcelona SuperComputing Center, and by the USQCD
collaboration.  Parts of these calculations were performed using the
{\tt chroma} lattice field theory library \cite{Edwards:2004sx}.

\bibliography{MM}

\begin{thebibliography}{27}
\expandafter\ifx\csname natexlab\endcsname\relax\def\natexlab#1{#1}\fi
\expandafter\ifx\csname bibnamefont\endcsname\relax
  \def\bibnamefont#1{#1}\fi
\expandafter\ifx\csname bibfnamefont\endcsname\relax
  \def\bibfnamefont#1{#1}\fi
\expandafter\ifx\csname citenamefont\endcsname\relax
  \def\citenamefont#1{#1}\fi
\expandafter\ifx\csname url\endcsname\relax
  \def\url#1{\texttt{#1}}\fi
\expandafter\ifx\csname urlprefix\endcsname\relax\def\urlprefix{URL }\fi
\providecommand{\bibinfo}[2]{#2}
\providecommand{\eprint}[2][]{\url{#2}}

\bibitem[{\citenamefont{Sheikholeslami and
  Wohlert}(1985)}]{Sheikholeslami:1985ij}
\bibinfo{author}{\bibfnamefont{B.}~\bibnamefont{Sheikholeslami}}
  \bibnamefont{and} \bibinfo{author}{\bibfnamefont{R.}~\bibnamefont{Wohlert}},
  \bibinfo{journal}{Nucl.Phys.} \textbf{\bibinfo{volume}{B259}},
  \bibinfo{pages}{572} (\bibinfo{year}{1985}).

\bibitem[{\citenamefont{L{\"u}scher and Weisz}(1985)}]{Luscher:1984xn}
\bibinfo{author}{\bibfnamefont{M.}~\bibnamefont{L{\"u}scher}} \bibnamefont{and}
  \bibinfo{author}{\bibfnamefont{P.}~\bibnamefont{Weisz}},
  \bibinfo{journal}{Commun.Math.Phys.} \textbf{\bibinfo{volume}{97}},
  \bibinfo{pages}{59} (\bibinfo{year}{1985}).

\bibitem[{\citenamefont{Detmold and Orginos}(2013)}]{Detmold:2012eu}
\bibinfo{author}{\bibfnamefont{W.}~\bibnamefont{Detmold}} \bibnamefont{and}
  \bibinfo{author}{\bibfnamefont{K.}~\bibnamefont{Orginos}},
  \bibinfo{journal}{Phys.Rev.} \textbf{\bibinfo{volume}{D87}},
  \bibinfo{pages}{114512} (\bibinfo{year}{2013}), \eprint{1207.1452}.

\bibitem[{\citenamefont{Beane et~al.}(2013{\natexlab{a}})\citenamefont{Beane,
  Chang, Cohen, Detmold, Lin et~al.}}]{Beane:2012vq}
\bibinfo{author}{\bibfnamefont{S.}~\bibnamefont{Beane}},
  \bibinfo{author}{\bibfnamefont{E.}~\bibnamefont{Chang}},
  \bibinfo{author}{\bibfnamefont{S.}~\bibnamefont{Cohen}},
  \bibinfo{author}{\bibfnamefont{W.}~\bibnamefont{Detmold}},
  \bibinfo{author}{\bibfnamefont{H.}~\bibnamefont{Lin}}, \bibnamefont{et~al.},
  \bibinfo{journal}{Phys.Rev.} \textbf{\bibinfo{volume}{D87}},
  \bibinfo{pages}{034506} (\bibinfo{year}{2013}{\natexlab{a}}),
  \eprint{1206.5219}.

\bibitem[{\citenamefont{Beane et~al.}(2013{\natexlab{b}})}]{Beane:2013br}
\bibinfo{author}{\bibfnamefont{S.}~\bibnamefont{Beane}} \bibnamefont{et~al.}
  (\bibinfo{collaboration}{NPLQCD Collaboration}), \bibinfo{journal}{Phys.Rev.}
  \textbf{\bibinfo{volume}{C88}}, \bibinfo{pages}{024003}
  (\bibinfo{year}{2013}{\natexlab{b}}), \eprint{1301.5790}.

\bibitem[{\citenamefont{Bernard et~al.}(1982)\citenamefont{Bernard, Draper,
  Olynyk, and Rushton}}]{Bernard:1982yu}
\bibinfo{author}{\bibfnamefont{C.~W.} \bibnamefont{Bernard}},
  \bibinfo{author}{\bibfnamefont{T.}~\bibnamefont{Draper}},
  \bibinfo{author}{\bibfnamefont{K.}~\bibnamefont{Olynyk}}, \bibnamefont{and}
  \bibinfo{author}{\bibfnamefont{M.}~\bibnamefont{Rushton}},
  \bibinfo{journal}{Phys.Rev.Lett.} \textbf{\bibinfo{volume}{49}},
  \bibinfo{pages}{1076} (\bibinfo{year}{1982}).

\bibitem[{\citenamefont{Martinelli et~al.}(1982)\citenamefont{Martinelli,
  Parisi, Petronzio, and Rapuano}}]{Martinelli:1982cb}
\bibinfo{author}{\bibfnamefont{G.}~\bibnamefont{Martinelli}},
  \bibinfo{author}{\bibfnamefont{G.}~\bibnamefont{Parisi}},
  \bibinfo{author}{\bibfnamefont{R.}~\bibnamefont{Petronzio}},
  \bibnamefont{and} \bibinfo{author}{\bibfnamefont{F.}~\bibnamefont{Rapuano}},
  \bibinfo{journal}{Phys.Lett.} \textbf{\bibinfo{volume}{B116}},
  \bibinfo{pages}{434} (\bibinfo{year}{1982}).

\bibitem[{\citenamefont{Lee et~al.}(2005)\citenamefont{Lee, Kelly, Zhou, and
  Wilcox}}]{Lee:2005ds}
\bibinfo{author}{\bibfnamefont{F.}~\bibnamefont{Lee}},
  \bibinfo{author}{\bibfnamefont{R.}~\bibnamefont{Kelly}},
  \bibinfo{author}{\bibfnamefont{L.}~\bibnamefont{Zhou}}, \bibnamefont{and}
  \bibinfo{author}{\bibfnamefont{W.}~\bibnamefont{Wilcox}},
  \bibinfo{journal}{Phys.Lett.} \textbf{\bibinfo{volume}{B627}},
  \bibinfo{pages}{71} (\bibinfo{year}{2005}), \eprint{hep-lat/0509067}.

\bibitem[{\citenamefont{Lee et~al.}(2006)\citenamefont{Lee, Zhou, Wilcox, and
  Christensen}}]{Lee:2005dq}
\bibinfo{author}{\bibfnamefont{F.~X.} \bibnamefont{Lee}},
  \bibinfo{author}{\bibfnamefont{L.}~\bibnamefont{Zhou}},
  \bibinfo{author}{\bibfnamefont{W.}~\bibnamefont{Wilcox}}, \bibnamefont{and}
  \bibinfo{author}{\bibfnamefont{J.~C.} \bibnamefont{Christensen}},
  \bibinfo{journal}{Phys.Rev.} \textbf{\bibinfo{volume}{D73}},
  \bibinfo{pages}{034503} (\bibinfo{year}{2006}), \eprint{hep-lat/0509065}.

\bibitem[{\citenamefont{Detmold et~al.}(2006)\citenamefont{Detmold, Tiburzi,
  and Walker-Loud}}]{Detmold:2006vu}
\bibinfo{author}{\bibfnamefont{W.}~\bibnamefont{Detmold}},
  \bibinfo{author}{\bibfnamefont{B.}~\bibnamefont{Tiburzi}}, \bibnamefont{and}
  \bibinfo{author}{\bibfnamefont{A.}~\bibnamefont{Walker-Loud}},
  \bibinfo{journal}{Phys.Rev.} \textbf{\bibinfo{volume}{D73}},
  \bibinfo{pages}{114505} (\bibinfo{year}{2006}), \eprint{hep-lat/0603026}.

\bibitem[{\citenamefont{Aubin et~al.}(2009)\citenamefont{Aubin, Orginos,
  Pascalutsa, and Vanderhaeghen}}]{Aubin:2008qp}
\bibinfo{author}{\bibfnamefont{C.}~\bibnamefont{Aubin}},
  \bibinfo{author}{\bibfnamefont{K.}~\bibnamefont{Orginos}},
  \bibinfo{author}{\bibfnamefont{V.}~\bibnamefont{Pascalutsa}},
  \bibnamefont{and}
  \bibinfo{author}{\bibfnamefont{M.}~\bibnamefont{Vanderhaeghen}},
  \bibinfo{journal}{Phys.Rev.} \textbf{\bibinfo{volume}{D79}},
  \bibinfo{pages}{051502} (\bibinfo{year}{2009}), \eprint{0811.2440}.

\bibitem[{\citenamefont{Detmold et~al.}(2009)\citenamefont{Detmold, Tiburzi,
  and Walker-Loud}}]{Detmold:2009dx}
\bibinfo{author}{\bibfnamefont{W.}~\bibnamefont{Detmold}},
  \bibinfo{author}{\bibfnamefont{B.~C.} \bibnamefont{Tiburzi}},
  \bibnamefont{and}
  \bibinfo{author}{\bibfnamefont{A.}~\bibnamefont{Walker-Loud}},
  \bibinfo{journal}{Phys.Rev.} \textbf{\bibinfo{volume}{D79}},
  \bibinfo{pages}{094505} (\bibinfo{year}{2009}), \eprint{0904.1586}.

\bibitem[{\citenamefont{Detmold et~al.}(2010)\citenamefont{Detmold, Tiburzi,
  and Walker-Loud}}]{Detmold:2010ts}
\bibinfo{author}{\bibfnamefont{W.}~\bibnamefont{Detmold}},
  \bibinfo{author}{\bibfnamefont{B.}~\bibnamefont{Tiburzi}}, \bibnamefont{and}
  \bibinfo{author}{\bibfnamefont{A.}~\bibnamefont{Walker-Loud}},
  \bibinfo{journal}{Phys.Rev.} \textbf{\bibinfo{volume}{D81}},
  \bibinfo{pages}{054502} (\bibinfo{year}{2010}), \eprint{1001.1131}.

\bibitem[{\citenamefont{Primer et~al.}(2014)\citenamefont{Primer, Kamleh,
  Leinweber, and Burkardt}}]{Primer:2013pva}
\bibinfo{author}{\bibfnamefont{T.}~\bibnamefont{Primer}},
  \bibinfo{author}{\bibfnamefont{W.}~\bibnamefont{Kamleh}},
  \bibinfo{author}{\bibfnamefont{D.}~\bibnamefont{Leinweber}},
  \bibnamefont{and} \bibinfo{author}{\bibfnamefont{M.}~\bibnamefont{Burkardt}},
  \bibinfo{journal}{Phys.Rev.} \textbf{\bibinfo{volume}{D89}},
  \bibinfo{pages}{034508} (\bibinfo{year}{2014}), \eprint{1307.1509}.

\bibitem[{\citenamefont{Fiebig et~al.}(1989)\citenamefont{Fiebig, Wilcox, and
  Woloshyn}}]{Fiebig:1988en}
\bibinfo{author}{\bibfnamefont{H.}~\bibnamefont{Fiebig}},
  \bibinfo{author}{\bibfnamefont{W.}~\bibnamefont{Wilcox}}, \bibnamefont{and}
  \bibinfo{author}{\bibfnamefont{R.}~\bibnamefont{Woloshyn}},
  \bibinfo{journal}{Nucl.Phys.} \textbf{\bibinfo{volume}{B324}},
  \bibinfo{pages}{47} (\bibinfo{year}{1989}).

\bibitem[{\citenamefont{Christensen et~al.}(2005)\citenamefont{Christensen,
  Wilcox, Lee, and Zhou}}]{Christensen:2004ca}
\bibinfo{author}{\bibfnamefont{J.~C.} \bibnamefont{Christensen}},
  \bibinfo{author}{\bibfnamefont{W.}~\bibnamefont{Wilcox}},
  \bibinfo{author}{\bibfnamefont{F.~X.} \bibnamefont{Lee}}, \bibnamefont{and}
  \bibinfo{author}{\bibfnamefont{L.-m.} \bibnamefont{Zhou}},
  \bibinfo{journal}{Phys.Rev.} \textbf{\bibinfo{volume}{D72}},
  \bibinfo{pages}{034503} (\bibinfo{year}{2005}), \eprint{hep-lat/0408024}.

\bibitem[{\citenamefont{Lujan et~al.}(2014)\citenamefont{Lujan, Alexandru,
  Freeman, and Lee}}]{Lujan:2014kia}
\bibinfo{author}{\bibfnamefont{M.}~\bibnamefont{Lujan}},
  \bibinfo{author}{\bibfnamefont{A.}~\bibnamefont{Alexandru}},
  \bibinfo{author}{\bibfnamefont{W.}~\bibnamefont{Freeman}}, \bibnamefont{and}
  \bibinfo{author}{\bibfnamefont{F.}~\bibnamefont{Lee}},
  \bibinfo{journal}{Phys.Rev.} \textbf{\bibinfo{volume}{D89}},
  \bibinfo{pages}{074506} (\bibinfo{year}{2014}), \eprint{1402.3025}.

\bibitem[{\citenamefont{'t~Hooft}(1979)}]{tHooft:1979uj}
\bibinfo{author}{\bibfnamefont{G.}~\bibnamefont{'t~Hooft}},
  \bibinfo{journal}{Nucl.Phys.} \textbf{\bibinfo{volume}{B153}},
  \bibinfo{pages}{141} (\bibinfo{year}{1979}).

\bibitem[{\citenamefont{Hill and Paz}(2011)}]{Hill:2011wy}
\bibinfo{author}{\bibfnamefont{R.~J.} \bibnamefont{Hill}} \bibnamefont{and}
  \bibinfo{author}{\bibfnamefont{G.}~\bibnamefont{Paz}},
  \bibinfo{journal}{Phys.Rev.Lett.} \textbf{\bibinfo{volume}{107}},
  \bibinfo{pages}{160402} (\bibinfo{year}{2011}), \eprint{1103.4617}.

\bibitem[{\citenamefont{Lee and Tiburzi}(2014)}]{Lee:2014iha}
\bibinfo{author}{\bibfnamefont{J.-W.} \bibnamefont{Lee}} \bibnamefont{and}
  \bibinfo{author}{\bibfnamefont{B.~C.} \bibnamefont{Tiburzi}}
  (\bibinfo{year}{2014}), \eprint{1407.8159}.

\bibitem[{\citenamefont{Chen et~al.}(1998)\citenamefont{Chen, Grie{\ss}hammer,
  Savage, and Springer}}]{Chen:1998vi}
\bibinfo{author}{\bibfnamefont{J.-W.} \bibnamefont{Chen}},
  \bibinfo{author}{\bibfnamefont{H.~W.} \bibnamefont{Grie{\ss}hammer}},
  \bibinfo{author}{\bibfnamefont{M.~J.} \bibnamefont{Savage}},
  \bibnamefont{and} \bibinfo{author}{\bibfnamefont{R.~P.}
  \bibnamefont{Springer}}, \bibinfo{journal}{Nucl.Phys.}
  \textbf{\bibinfo{volume}{A644}}, \bibinfo{pages}{221} (\bibinfo{year}{1998}),
  \eprint{nucl-th/9806080}.

\bibitem[{\citenamefont{Schmidt}(1937)}]{schmidt:1937}
\bibinfo{author}{\bibfnamefont{T.}~\bibnamefont{Schmidt}}, \bibinfo{journal}{Z.
  Phys.} \textbf{\bibinfo{volume}{106}}, \bibinfo{pages}{358}
  (\bibinfo{year}{1937}).

\bibitem[{\citenamefont{Foldy and Milford}(1950)}]{foldy:1950}
\bibinfo{author}{\bibfnamefont{L.~L.} \bibnamefont{Foldy}} \bibnamefont{and}
  \bibinfo{author}{\bibfnamefont{F.~J.} \bibnamefont{Milford}},
  \bibinfo{journal}{Phys.Rev,} \textbf{\bibinfo{volume}{80}},
  \bibinfo{pages}{751} (\bibinfo{year}{1950}).

\bibitem[{\citenamefont{Walker-Loud}(2013)}]{Walker-Loud:2013yua}
\bibinfo{author}{\bibfnamefont{A.}~\bibnamefont{Walker-Loud}},
  \bibinfo{journal}{PoS} \textbf{\bibinfo{volume}{CD12}}, \bibinfo{pages}{017}
  (\bibinfo{year}{2013}), \eprint{1304.6341}.

\bibitem[{\citenamefont{Piarulli et~al.}(2013)\citenamefont{Piarulli, Girlanda,
  Marcucci, Pastore, Schiavilla et~al.}}]{Piarulli:2012bn}
\bibinfo{author}{\bibfnamefont{M.}~\bibnamefont{Piarulli}},
  \bibinfo{author}{\bibfnamefont{L.}~\bibnamefont{Girlanda}},
  \bibinfo{author}{\bibfnamefont{L.}~\bibnamefont{Marcucci}},
  \bibinfo{author}{\bibfnamefont{S.}~\bibnamefont{Pastore}},
  \bibinfo{author}{\bibfnamefont{R.}~\bibnamefont{Schiavilla}},
  \bibnamefont{et~al.}, \bibinfo{journal}{Phys.Rev.}
  \textbf{\bibinfo{volume}{C87}}, \bibinfo{pages}{014006}
  (\bibinfo{year}{2013}), \eprint{1212.1105}.

\bibitem[{\citenamefont{Tiburzi}(2014)}]{Tiburzi:2014zva}
\bibinfo{author}{\bibfnamefont{B.~C.} \bibnamefont{Tiburzi}},
  \bibinfo{journal}{Phys.Rev.} \textbf{\bibinfo{volume}{D89}},
  \bibinfo{pages}{074019} (\bibinfo{year}{2014}), \eprint{1403.0878}.

\bibitem[{\citenamefont{Edwards and Joo}(2005)}]{Edwards:2004sx}
\bibinfo{author}{\bibfnamefont{R.~G.} \bibnamefont{Edwards}} \bibnamefont{and}
  \bibinfo{author}{\bibfnamefont{B.}~\bibnamefont{Joo}},
  \bibinfo{journal}{Nucl.Phys.Proc.Suppl.} \textbf{\bibinfo{volume}{140}},
  \bibinfo{pages}{832} (\bibinfo{year}{2005}), \eprint{hep-lat/0409003}.

\end{thebibliography}

\end{document}